\documentstyle[12pt]{article}
\title{The Ward identity and nonadiabatic corrections to the quasiparticle
self-energy}
\author{O. V. Danylenko, O. V. Dolgov, V. V. Losyakov \\
I. E. Tamm Theoretical Department, \\
P. N. Lebedev Physical Institute, \\
Leninsky pr., 53, 117924, Moscow, Russia} \date{} 
\begin{document}
\maketitle
\begin{abstract}

\label{abstractpage}
In some newly discovered materials the ratio of phonon to electron energies
is no longer small. We have investigated the basement of
the recently proposed gauge-invariant self-consistent
method and found conditions of its applicability.
\end{abstract}

%

\label{introductionpage}
When describing electron-phonon interaction in usual metals, so-called
adiabatic approximation is used. It assumes that due to electroneutrality fast
electrons must follow "slow" ions. Narrow bandwidths and strong electron-phonon
interaction make the orders of these velocities comparable. A question appears:
Does this result in a breakdown of adiabaticity and increase of interelectron
interaction (and in the result $T_c$)?

Traditionally such corrections were considered to be small as the Migdal
parameter $\Omega _{ph}/ W$ was small \cite{M,ES,S,AM}. But recently there
were discovered materials (for example, fullerenes and high $T_c$
superconductors) where $\Omega _{ph} \sim W$. So Migdal's theorem about
smallness of adiabatic corrections \cite{M,AM} is violated. This gives the
grounds for not just discussing these corrections, but even for
considering an antiadiabatic limit $\Omega _{ph} \gg W$. For example,
it is
stated that this results in increase of interelectron interaction
and can increase $T_c$ for superconductors \cite{PSG1}. But even with small
$ \Omega_{ph}/ W$ there are contradictory conclusions about importance of
such corrections (the author of Ref. \cite{M} considers them to be negligible,
in contradiction to the authors of Refs. \cite{PSG1,T1}).

There are two methods to take nonadiabatic corrections into account. One of
them, which can be called Migdal's one, is based on the solution to the
Bethe-Salpeter equation for the vertex function \cite{M,ES,S,DDLLT}. In the lowest
approximation, the first correction to the unit vertex is determined by the
diagram in \mbox{Fig. 1 (a)}. Then with the found vertex function the self-energy
$\Sigma$ is calculated (see \mbox{Fig. 1 (b)}). This traditional method does not
allow to take into account higher order diagrams analytically because of
complexity of integration over momenta.

Recently Y.~Takada offered another method \cite{T1} which he called
the gauge-invariant self-consistent (GISC). In this method, based on the Ward
identity \cite{ES}, the vertex function is chosen as a functional of the
self-energy which is supposed to be independent from momentum. So, when
comparing this method with Migdal's one, we discuss validity of such
approach. The parameter
$\Omega_{ph}/ W$ is considered to be small.

\label{modelpage}
For simplicity in the Letter we consider the model of the Fermi liquid with the
usual
Fr\"ohlich Hamiltonian \cite{AM} and the Einstein phonon spectrum. This
corresponds to $\alpha ^2F_E(\Omega )=\frac12 \lambda \Omega _{ph} \delta
(\Omega -\Omega _{ph})$ where $\Omega _{ph}$ is the phonon frequency
independent from momentum, and $ \lambda $ is the mass-enhancement parameter.
It is assumed that the density of states is the constant $N(0)$ for
$-W<\varepsilon <W$ where $2W=2p_Fv_F$ is the bandwidth. The chemical potential
equals zero which corresponds to a half-filled band.

We are interested in the quasiparticle self-energy $\Sigma $ on the imaginary
axis which is determined by the diagram in \mbox{Fig. 1 (b)}. The phonon Green's
function $D_0(i\omega _\nu )=- \Omega _{ph}^2 (\omega _\nu ^2+\Omega
_{ph}^2)^{-1}$, $\omega _\nu =2\nu \pi T$, and the electron Green's function
$G(i\omega _n,\vec p)=(i\omega _n-\varepsilon _{\vec
p}-\Sigma (i\omega _n,\vec p))^{-1}$, $\omega _n=(2n+1)\pi T $, $\varepsilon
_{\vec p}$ is the electron spectrum.

Assuming that $\Sigma (i\omega _n,\vec k)$ weakly depends on
$\vec k$, we take $| \vec k| =p_F=const$, and so
$\Sigma =\Sigma (i\omega _n)$. Then the diagram in \mbox{Fig. 1 (b)}
corresponds to \cite{AM}
\begin{eqnarray}
\Sigma (i\omega _n)=\int\limits_0^\infty d\Omega
\alpha ^2F_E(\Omega )T\sum\limits_{\omega _\nu }\frac{2\Omega }{\omega _\nu
^2+\Omega ^2} \frac 1{N(0)} \nonumber \\ \times \sum\limits_{\vec q}G(i\omega
_n-i\omega _\nu ,\vec k-\vec q)\Gamma (i\omega _n,i\omega _n-i\omega _\nu ,\vec
q) \label{eq1}
\end{eqnarray}

Takada's and Migdal's methods differ by a choice of
the vertex function, and this is discussed below. We will obtain the
results for $T=0$ and the variables will change as follows:
$\omega _n\longrightarrow \omega $, $\omega _\nu
\longrightarrow \omega ^{\prime }$ remaining on the imaginary axis.

\label{takadapage}
According to Takada's GISC method, the vertex function can be chosen as a
functional of the self-energy \cite{T1} $\Gamma_T=\Gamma _T[\Sigma _T]$.  This
choice is based on the Ward identity \cite{ES}
\begin{eqnarray}
i\omega _\nu
\Gamma (i\omega _n,i\omega _n-i\omega _\nu ,\vec k,\vec k-\vec q)-\vec q\vec
\Gamma
(i\omega _n,i\omega _n-i\omega _\nu ,\vec k,\vec k-\vec q) \nonumber \\
=G^{-1}(i\omega _n,\vec k)-G^{-1}(i\omega _n-i\omega _\nu ,\vec k-\vec q),
\label{eq1.5}
\end{eqnarray}
valid for all $\omega_\nu $ and $\vec q $.
It is essential that the phonon Green's function does not depend on momentum.
Only in this case one may use the Ward identity for finite momentum
$\vec q $ \cite{ES}.
In Ref. \cite{T1} Takada proposes to use this identity for
$v_F| \vec q| \ll | \omega _\nu | $ and choose
\cite{N}
\begin{eqnarray}
\Gamma _T[\Sigma ]=\Gamma _T(i\omega _n,i\omega
_n-i\omega _\nu )=1+(- \frac{\Sigma _T(i\omega _n)}{2i\omega _n}-\frac{\Sigma
_T(i\omega _n-i\omega _\nu )}{2(i\omega _n-i\omega _\nu )}), \label{eq2}
\end{eqnarray}
which corresponds to the estimates of Refs. \cite{ES,S} in this limit.

The set of equations (\ref{eq1}), (\ref{eq2}) is proposed to be solved
by iterations.  At the first step $G=G^{(0)}$,
$\Gamma =\Gamma _T^{(1)}=1$ give $\Sigma _T=\Sigma _T^{(1)}$ (Fig. 2 (a)).
For $T=0$
\begin{eqnarray}
\Sigma _T^{(1)}(i\omega )=-i\lambda \Omega
_{ph}\arctan \frac \omega {\Omega _{ph}} + i\lambda \Omega _{ph}\arctan \frac
\omega {W+\Omega _{ph}},
\label{eq4.5}
\end{eqnarray}
which corresponds to the results of Refs. \cite{ES,PSG1}. At the second step
$G=G[\Sigma _T^{(1)}]$, and the correction $\Gamma _T^{(2)}$ for
$\Omega_{ph}/W \ll 1$ gives:
\begin{eqnarray}
\Sigma _T^{(2)}(i\omega )=i\lambda ^2\Omega _{ph}b(m),
\label{eqB4} \mbox{ where $m= \omega / \Omega _{ph}$, $b(m)=b_1(m)+b_2(m)$.
}
\end{eqnarray}

\begin{eqnarray}
b_1(m)=
-\frac{1}{2m} \arctan ^2m , \label{eqB5}
\end{eqnarray}

\begin{eqnarray}
b_2(m)=\frac 14\int\limits_{-\infty }^{+\infty }dy\frac 1{y^2+1}\frac{
\arctan (m-y)}{m-y}{\mbox{sign}}(y-m),
\label{eqB6} \mbox{ $y= \omega ^{\prime } / \Omega_{ph}$.
}
\end{eqnarray}

So
\begin{eqnarray}
b(m)=-\left\{
\begin{array}{r}
(\frac 14+\frac{\pi ^2}{16})m, \mbox{ } |m|\ll 1, \\
\frac{\pi ^2}4\frac 1m, \mbox{ } |m|\gg 1
\end{array}
\right. \ .
\label{eqB10}
\end{eqnarray}

Here $\Sigma^{(2)}_T$ is obtained by using (\ref{eq4.5})
with infinite limits of integration over $d \varepsilon$.
During this calculation the vertex $\Gamma^{(2)}_T$ is of the order of $\lambda$
for any $q$ which is incorrect. So in Takada's method the leading term of the
corrections does not have the small parameter $\Omega _{ph}/ W$, that is in
this approach Migdal's theorem does not hold for the self-energy. This can be
explained by the fact that the region $v_F| \vec q| \ll | \omega ^{\prime}| $
where $\Gamma \sim
\lambda $ gives in reality a small contribution to $\Sigma $. If, on the
contrary, $v_F| \vec q|
\gg | \omega ^{\prime }| $, the vertex $ \Gamma \sim \lambda \Omega _{ph}/ W$
which gives $ \Sigma^{(2)}\sim \lambda \Omega _{ph}^2/ W$.
At the same time, the sign of the vertex correction to
the self-energy in GISC
method is opposite to that in Migdal's one (see below).

\label{migdalpage}
In the method which can be called Migdal's one, there is considered the first
correction to the unit vertex function $\Gamma ^{(1)}$. It is shown on
\mbox{Fig. 1 (a)}. It has been estimated in many papers \cite{M,S},
but now it is becoming especially significant \cite{PSG1}
as it has appeared that the parameter
$\lambda \Omega_{ph}/ W$ can be not very small.

The diagram in Fig. 1 (a) corresponds to the expression
\begin{eqnarray}
\Gamma ^{(2)}(i\omega _n,i\omega _n-i\omega _\nu ,\vec q)
=
T\sum\limits_{\omega
_{\nu ^{\prime }}}\sum\limits_{\vec q\mbox{ }^{\prime }}\int\limits_0^\infty
d\Omega
\frac{\alpha ^2F_E(\Omega )2\Omega }{N(0)(\omega _{\nu ^{\prime }}^2+\Omega
^2)} \nonumber \\
\times G^{(0)}(i\omega _n-i\omega _{\nu ^{\prime }},\vec p-\vec q\mbox{
}^{\prime }) G^{(0)}(i\omega _n-i\omega _{\nu ^{\prime }}-i\omega _\nu ,\vec
p-\vec q\mbox{ }^{\prime }-\vec q).  \label{eq3.5} \end{eqnarray}

Assuming that $| $ $\vec p-\vec q\mbox{ }^{\prime }| \sim p_F$ and expanding
$\varepsilon
(\vec p-\vec q\mbox{ }^{\prime }-\vec q)\approx \varepsilon (\vec p-\vec
q\mbox{ }^{\prime
})-qv_F\cos \theta $, we get for $T=0$:
\begin{eqnarray}
\Gamma ^{(2)}(i\omega ,i\omega -i\omega ^{\prime },\vec q)=\lambda
\frac{\Omega _{ph}}{qv_F}\arctan \frac{qv_F}{\omega ^{\prime }}
(
\arctan\frac \omega {\Omega _{ph}}-\arctan \frac{\omega -\omega ^{\prime }}{\Omega
_{ph}} \nonumber \\
-\arctan\frac \omega {W+\Omega _{ph}}+\arctan \frac{\omega -\omega ^{\prime
}}{W+\Omega _{ph}}
).
\label{eq4}
\end{eqnarray}
At $v_F| \vec q| \ll | \omega ^{\prime }| $ this expression corresponds to the
Ward identity (\ref{eq1.5}) and to the choice of a functional (\ref{eq2}).
For $v_F| \vec q| \gg | \omega ^{\prime }| $ the correction
$\Gamma ^{(2)}(i\omega ,i\omega -i\omega ^{\prime },\vec q)\sim \lambda
\Omega _{ph}/ qv_F \sim \lambda \Omega _{ph}/ W
$, that is Migdal's theorem for the vertex function does hold in this method.

The corresponding self-energy then equals (see Fig. 2)
$$\Sigma=\Sigma^{(1)}+\Sigma_v^{(2)}. $$
$\Sigma^{(1)}$ is equal to
$\Sigma _T^{(1)}$ as it also corresponds to $\Gamma ^{(1)}=1$.
For the correction due to $\Gamma^{(2)}$, $\Sigma _v^{(2)}$, one can get at
$\Omega_{ph}/W \ll 1$:

$$\Sigma _v^{(2)}(i\omega )=i\lambda ^2\frac{\Omega _{ph}^2}{W}d_v(m)
 \mbox{ where }m=\frac{\omega}{\Omega_{ph}},$$

\begin{eqnarray}
d_v(m)=\frac \pi 4\int\limits_{-\infty }^\infty dy\frac 1{y^2+1}{\mbox{sign}}(y){
\mbox{sign}}
(y-m)(\arctan m-\arctan (m-y)).
\label{eqd05}
\end{eqnarray}

In the limiting cases
\begin{eqnarray}
d_v(m)=\left\{
\begin{array}{r}
\frac{\pi ^2}{8}m, \mbox{ } |m|\ll 1, \\
\frac{\pi ^2}4\frac 1m, \mbox{ } |m|\gg 1
\end{array}
\right. \ .
\label{eqd06}
\end{eqnarray}
The sign of the correction $\Sigma^{(2)}_v$ is opposite to that of
$\Sigma^{(1)}$ which favour tendency towards instability \cite{DDLLT,DDLKSF}.

In this approximation one can find the vector vertex function $\vec \Gamma $
and show that it really cannot be neglected. Using electron-hole symmetry, in our
model,
$\vec \Gamma(i\omega ,i\omega -i\omega ^{\prime },\vec k,\vec k-\vec q)
 =\vec \Gamma ^{(1)}+\vec \Gamma ^{(2)} \mbox{ where }
\vec \Gamma ^{(1)} \approx \frac{\vec k}{m},$
$$\vec q\vec \Gamma ^{(2)}
=-i\lambda \Omega _{ph}(1-\frac{\omega ^{\prime }}{qv_F}\arctan
\frac{qv_F}{\omega ^{\prime }})
(\arctan \frac \omega {\Omega _{ph}}-\arctan
\frac{\omega -\omega ^{\prime }}{\Omega _{ph}}).$$

The diagram in Fig. 1 (a) corresponds to $\vec \Gamma ^{(2)}$ if one substitutes
$\Gamma ^{(1)}$
by $\vec \Gamma ^{(1)}$ and $\Gamma ^{(2)}(i\omega _n,
i\omega_n
- i\omega _\nu, \vec q)$ - by $\vec \Gamma ^{(2)}(i\omega, i\omega -i\omega
^{\prime }, -\vec q)$.

If one designates $x=\omega ^{\prime } / qv_F$, then using at $\Omega_{ph}/W
\ll 1$ the expression (\ref{eq4}) for $\Gamma ^{(2)}$, we can find that the terms
$\omega ^{\prime }\Gamma ^{(2)}$ and $\vec q\vec \Gamma
^{(2)}$ from (\ref{eq1.5}) become of the same magnitude if $x_0\arctan (1/
x_0)=1-x_0\arctan (1/x_0)$, from which $x_0=0.43$. If, however, $\omega ^{\prime
}/
qv_F<x_0$, the vector term dominates and it may not be neglected.

\label{resultspage}
So in the Letter there is consistently considered contribution of
nonadiabatic effects to the electron self-energy in first orders in $\lambda$.
The two methods of taking into account the nonadiabatic corrections were compared:
1) Migdal's and 2) the GISC proposed by Takada.

It is shown that the latter gives for the self-energy physically incorrect result.
This is connected with the fact that the leading contribution to $\Sigma$
is made by the region $qv_F \gg \omega^{\prime}$ where $\Gamma \sim
\lambda \Omega _{ph}/ W$. This means that in the Ward identity one may not
neglect the vector term as proposed by Takada. In the frame of the standard
(Migdal's) approach, in the lowest order, there is evaluated the parameter
$x=\omega^{\prime}/qv_F$ at which contributions of the vector and scalar terms
become the same (see Fig. 3).
They are equal at $x_0=0.43$.
As it was shown above, the region of large q's indeed gives the main
contribution to the self-energy. That is why the GISC method, which
works only for small q's, may not be applied.

This work was supported by grants RFBR-96-02-16661a, INTAS-93-2154, and by
ISI Foundation and EU INTAS Network 1010-CT 930055.
The authors are also grateful to Prof. E.~G.~Maksimov for
useful discussions.



\thispagestyle{empty}
\vbox{
\centerline {\Large Figure Captures}
\label{figurespage}
\centerline { }

Fig. 1. (a) The equation for the vertex function $\Gamma^{(2)}$ in Migdal's
method.  The outgoing lines are shown for clearness and are not included in
the definition of $\Gamma$. (b) The equation for the self-energy $\Sigma$. The
wavy line is a bare phonon Green's function $D_0$, the double solid line is an
electron Green's function $G$, and the hatched triangle is a vertex function
$\Gamma$.

\centerline { }

Fig. 2 (a), (b). Diagrams which are included when calculating the self-energy
$\Sigma$. The solid straight line is a bare electron Green's function $G^{(0)}$.

\centerline { }

Fig. 3. The ratio of the vector to the scalar terms in the Ward identity. The
GISC method may only be used for large x's (small q's).
}
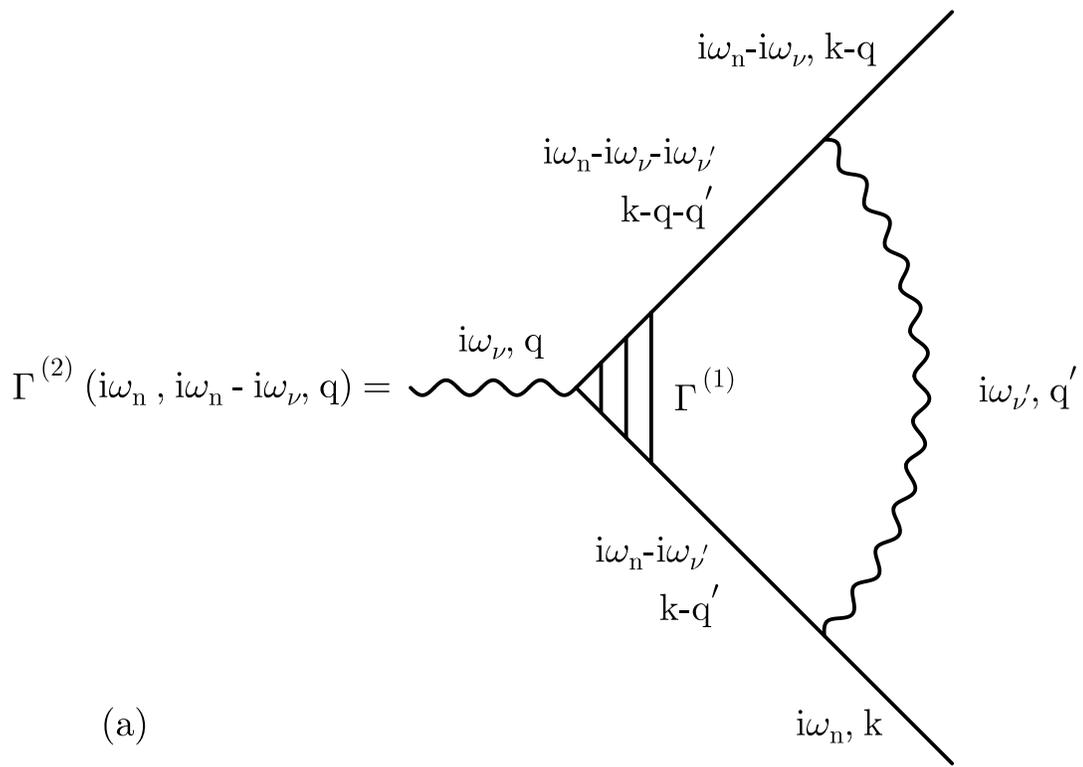
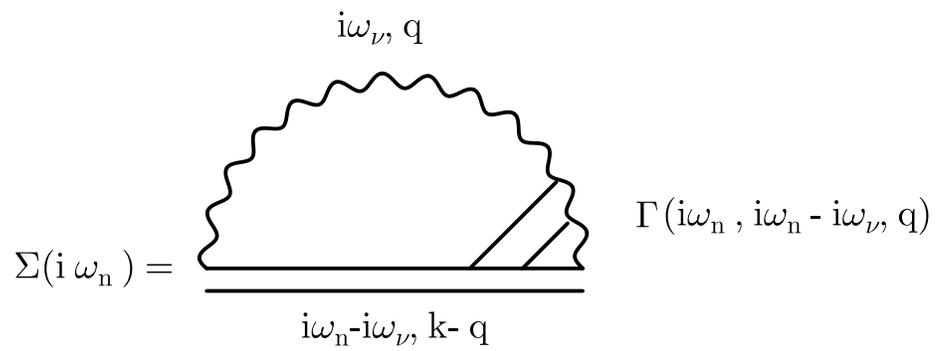
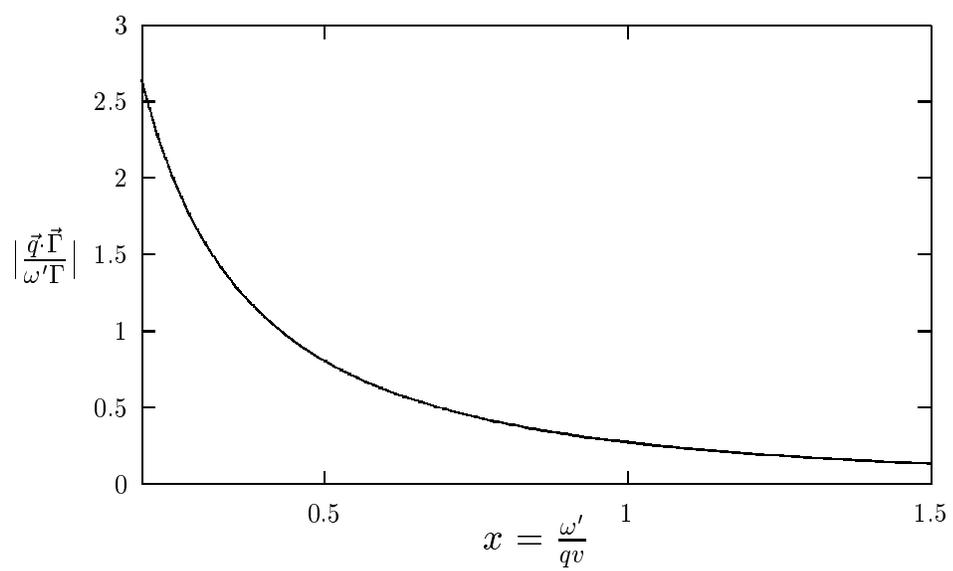
\end{document}